\documentclass[preprint,12pt]{elsarticle}

\usepackage{epsfig}

\usepackage{amssymb}

\journal{Astroparticle Physics}

\begin{document}

\begin{frontmatter}

\title{New Lower Limits on the Lifetime of Heavy Neutrino Radiative
Decay}

\author[unibo,infn,inaf]{S. Cecchini}
\author[unibo]{D. Centomo}
\author[unibo,infn]{G. Giacomelli}
\author[infn]{R. Giacomelli}
\author[unibo,infn]{M.~Giorgini}
\author[infn]{L. Patrizii}
\author[iss]{V. Popa\fnref{corr}}
\ead{vpopa@spacescience.ro}
\fntext[corr]{Corresponding author}
\author[unito,infnto]{C.G. \c{S}erb\u{a}nu\c{t}}
\address[unibo]{Dipartimento di Fisica dell'Universit\`{a}
 di Bologna, I-40127, Bologna, Italy}
 \address[infn]{INFN Sezione di Bologna, I-40127, Bologna, Italy}
\address[inaf]{IASF/INAF, I-40129 Bologna, Italy}
\address[iss]{Institutul de \c{S}tiin\c{t}e Spa\c{t}iale, R-77125,
Bucharest-M\u{a}gurele, Romania}
\address[unito]{Dipartimento di Fisica Generale dell'Universit\`{a} di
Torino, I-10125, Torino, Italy}
\address[infnto]{INFN Sezione di Torino, I-10125, Torino, Italy}

\begin{abstract}
{\small
The data collected during the 2006 total solar eclipse are analyzed in
the search for signals
produced by a hypothetical radiative decay of massive neutrinos. In the
absence of the expected
light pattern, we set lower limits for the massive neutrino
components proper lifetime. The reached sensitivity indicates that
these are the best limits obtainable with this method.
}
\end{abstract}

\begin{keyword}

Solar neutrinos \sep Radiative decays of massive neutrinos \sep Neutrino mass
and mixing \sep Total solar eclipses \sep Image processing

\PACS 96.60.Vg \sep 13.35. Hb \sep 14.60 Pq. \sep 95.85.Ry \sep 95.75.Mn

\end{keyword}

\end{frontmatter}

\section{Introduction}
\label{int}

The evidence for solar and atmospheric neutrino
oscillations %\cite{sno,soudan,macro,sk,lbl}
%\cite{sno}-\cite{lbl}
[1-5] implies that neutrinos have
non-vanishing masses, and that neutrino flavour eigenstates are
superpositions of mass eigenstates. D.W. Sciama pointed out the
possibility of $\nu$ radiative decay and the observational and theoretical
consequences \cite{sci}. Massive neutrinos could
undergo radiative decays; a possible decay mode is
$\nu_2 \rightarrow \nu_1 + \gamma$.

The neutrino radiative decay requires a non-vanishing neutrino magnetic
moment; stringent existing limits ($\mu_\nu < 0.9 \cdot 10^{-10} \mu_B$,
\cite{pdg}) apply to neutrino flavour eigenstates and cannot be directly
extended to dipole magnetic moments of neutrino mass eigenstates.

``Semi-indirect" limits on neutrino radiative decay have been obtained from
solar and atmospheric neutrino data. The current
interpretation of existing observations is that of neutrino
flavour oscillations, but the contribution from neutrino decays as a secondary
effect cannot be excluded. From the SNO solar neutrino
data, in ref. \cite{bandy03} a lower limit of
$\tau_0/m > 8.7 \cdot 10^{-5}$ s/eV was inferred ($\tau_0$ and $m$ are
the proper lifetime and mass of a decaying neutrino, respectively). By
combining available solar neutrino data, limits of
$\tau_0/m > 2.8 \cdot 10^{-5}$ s/eV \cite{jos02}, or, following different
assumptions, $\tau_0/m > 10^{-4}$ s/eV \cite{beacon02} were obtained.

Direct searches for radiative (anti)-neutrino decays were performed in the
vicinity of nuclear reactors (e.g. \cite{bouchez88}, yielding limits between
$\tau_0/m > 10^{-8}$ s/eV and $\tau_0/m > 0.1$ s/eV, for $\Delta m / m$
between $10^{-7}$ and 0.1); the Borexino Counting Test Facility at
Gran Sasso yielded limits at the level of $\tau_0/m \sim 10^3$  s/eV
\cite{derbin02}.

Cowsik \cite{cowsik77} pointed out that astronomical
observations at x-ray, optical and radio-frequencies could be used to derive
bounds on radiative lifetimes on the basis of the non-observation of the
final $\gamma$-ray. More recently bounds have been deduced from
infra-red background measurements. Such lower limits are large (e.g.
$\tau_0/m > 2.8 \cdot 10^{15}$ s/eV \cite{bludman92}), but they
are indirect and rather speculative.

The Sun is a strong source of $\nu_e$ neutrinos; the expected flux at the
Earth (neglecting oscillation effects) is $\Phi \simeq 7 \cdot 10^{10}$
cm$^{-2}$s$^{-1}$. If radiative neutrino decays occur yielding
visible photons, they would not be observable due to the very large amount of
solar light. During a Total Solar Eclipse (TSE) the Moon screens
the direct light from the Sun, while it is completely transparent to solar
neutrinos. An experiment looking for such an effect would thus be sensitive
to neutrino decays occurring between the Moon and the Earth.

In a pioneering experiment performed in occasion of the October 24, 1995 TSE,
a first search was made for visible photons emitted through radiative decays of
solar neutrinos during their flight between the Moon and the Earth
\cite{birnbaum97}. The authors assumed that all neutrinos have masses of the
order of few eV, $\Delta m^2_{12} \simeq 10^{-5}$ eV$^2$, an
energy of 860 keV and that all decays yield visible
photons, which travel nearly in the same direction as the parent
neutrinos. From the absence of a positive signal they estimated a lower limit
on $\tau_0$ (97 s) which, in view of the assumptions made, is now not reliable.

We made a search for solar neutrino radiative decays during the June
21, 2002 TSE, in Zambia \cite{noi} after a first attempt during the 1999
eclipse in Romania \cite{vechi}. The proper lower time limits (95\% C.L.)
obtained for the $\nu_2 \rightarrow \nu_1 + \gamma$  decays of left-handed
neutrinos ranged from $\tau_0/m_2 \simeq 10$ to $\simeq 10^9$ s/eV, for
$10^{-3}$ eV $ < m_1 < 0.1$ eV.

In this paper we report on the observations performed in occasion of the  29
March 2006 TSE, from Waw an Namos, Libya.

\section{The experimental setup}
\label{exp}

\subsection{The main observation system}
\label{main}

Our main system used a Matsukov-Cassegrain telescope ($\Phi=235$ mm,
$f = 2350$ mm) equipped with a fast 16 bit Mx916 CCD camera. The original
findscope was substituted by the digital videocamera used for
data taking during the 2002 TSE.

\begin{figure}
  \begin{center}
  \includegraphics[width=9cm]{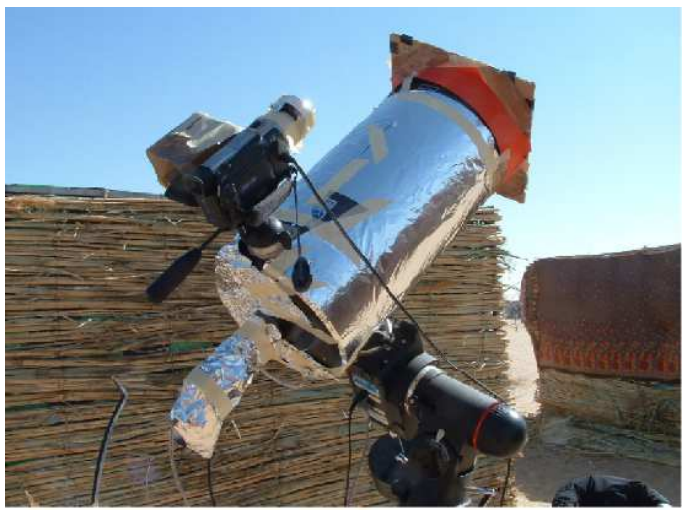}\\
  \caption{A photo of the main system deployed in the Libyan Sahara desert.
 The telescope, the CCD and the digital videocamera are visible. The aluminum
foil reduced the heating of the apparatus. The solar filter on the top of
the telescope was used only during the partiality. }\label{foto}
  \end{center}
\end{figure}

The night before the eclipse we aligned the system, adjusted the focus and took
calibration images of some standard luminosity stars (SAO99215 and SAO99802).
In order to avoid the
over-heating of the telescope and CCD and to minimize
the possibility of focus and alignment changes, the equipment was protected
with aluminum foils. A photograph of the set up is shown in Fig. \ref{foto}.
 Pictures of our set ups as well as data
evaluation may be found in ref. \cite{lib}.

The telescope movement was set to follow the Sun in order to have always the
center of the acquired images coincident with the Sun center. Furthermore, we
implemented a special CCD exposure algorithm in order to adapt the
exposure times to the luminosity level of the Moon image. The ashen
light (the Sun's light reflected by the Earth back to the Moon)
is one of the main background sources in such searches, but it allows
the reconstruction, frame
by frame, of the real position of the Sun behind the Moon, eliminating the
risk of pointing
errors due to undesired movements of the telescope.

\subsection{The backup systems}

The digital video-camera used in occasion of the 2002 TSE
was a small backup system. It produced a digital film of the
eclipse that could confirm our earlier results \cite{noi}.

A smaller Celestron C5 telescope equipped with a manually
controlled digital camera (Canon D20) was also used. We obtained 50
digital pictures of the eclipse.

In this paper we discuss only the data collected with the main system, which
has a much higher sensitivity than the back-up systems.

\section{The experimental data}
\label{dat}

 On the $29^{th}$ of March 2006 we observed the total solar eclipse
from a location
($17.960^\circ$ East longitude, $24.496^\circ$ North latitude and 465 m
altitude) in the Libyan Sahara desert, practically on the
totality line and very close to the maximum eclipse point. The time of the
totality was close to noon, so the Sun and the
Moon were near their highest positions in the sky (about $65.5^\circ$),
 corresponding to the minimum light absorption by the atmosphere.

The data collected by the main system consist in 212 digital
pictures of the central part of the ``dark" disk of the Moon. We used a
$2 \times 2$ binning in order to maximize the CCD sensitivity keeping at
the same time a good spatial resolution. Each image pixel covers
(in arc second squared) a solid angle of $ 1.99" \times 1.95"$.
The analysis described in Sect. \ref{ana} is based on the wavelet
decomposition of the images; we selected only the central largest dyadic
square from each frame ($256 \times 256$ square pixels); the total coverage
was $8.49' \times 8.32'$ (the Moon apparent diameter is $31'$).

\begin{figure}
  \begin{center}
  \includegraphics[width=9cm]{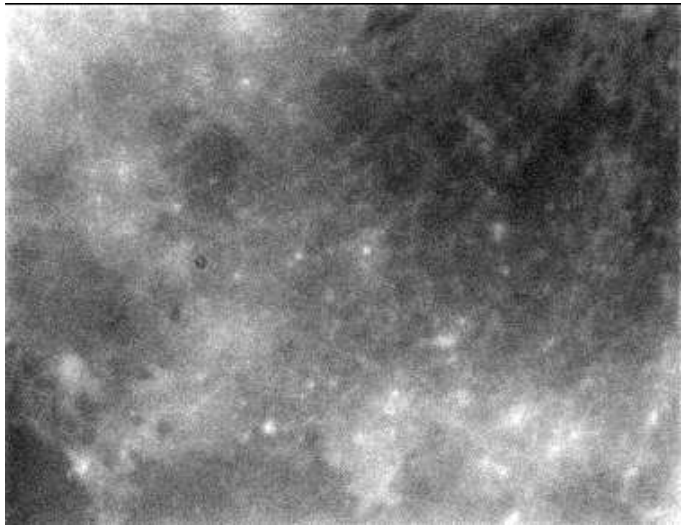}\\
  \caption{One of the frames recorded during the 2006 TSE. The center of the
image corresponds to the position of the center of the Sun behind the Moon.}
\label{frame}
  \end{center}
\end{figure}

Fig. \ref{frame} shows one of the frames recorded during the totality phase;
 details on the Moon surface seen in the ashen light are visible.

\begin{figure}
  \begin{center}
  \includegraphics[width=9cm]{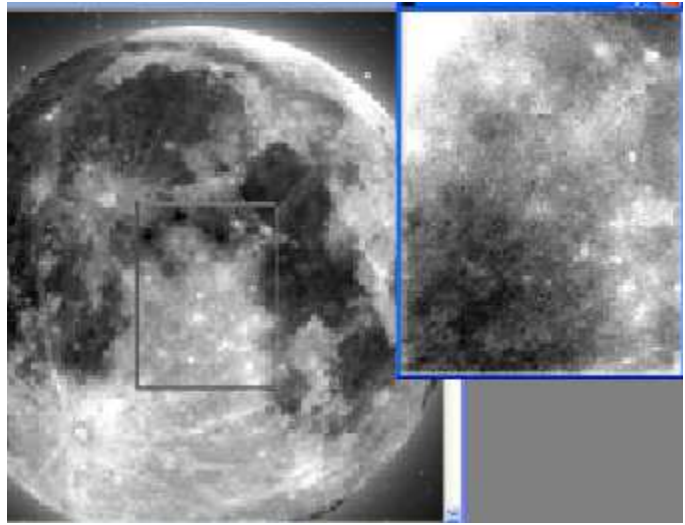}\\
  \caption{A full Moon picture (left) compared to one frame (right).
The differences are due to the different lighting conditions: in the light
 of the Sun and in the ashen light, respectively. The rectangle on the Moon
shows the area covered by the frame measured during the eclipse.}
\label{truc}
  \end{center}
\end{figure}

In Fig. \ref{truc} the image of the Moon observed in the light of the Sun
is compared with one of our frames (in ashen light). The corresponding area
on the Moon is marked by a rectangle. Differences are due to the different
Moon lighting conditions; some small scale details can be located in both
images.

Dark CCD frames, used to eliminate the effect of some possible ``hot pixels",
were recorded before and after the totality phase.

\subsection{Alignment and pointing}
\label{map}

The telescope was aligned the night before the eclipse using the North Star
(in the same observation session in which the calibration pictures were
taken), and it was kept in position till the end of the eclipse.

We continuously checked the Sun position, starting at the moment
of the first contact, using a grid applied on
the screen of the digital videocamera.

\begin{figure}
\vspace{-9cm}
  \begin{center}
  \includegraphics[width=15cm]{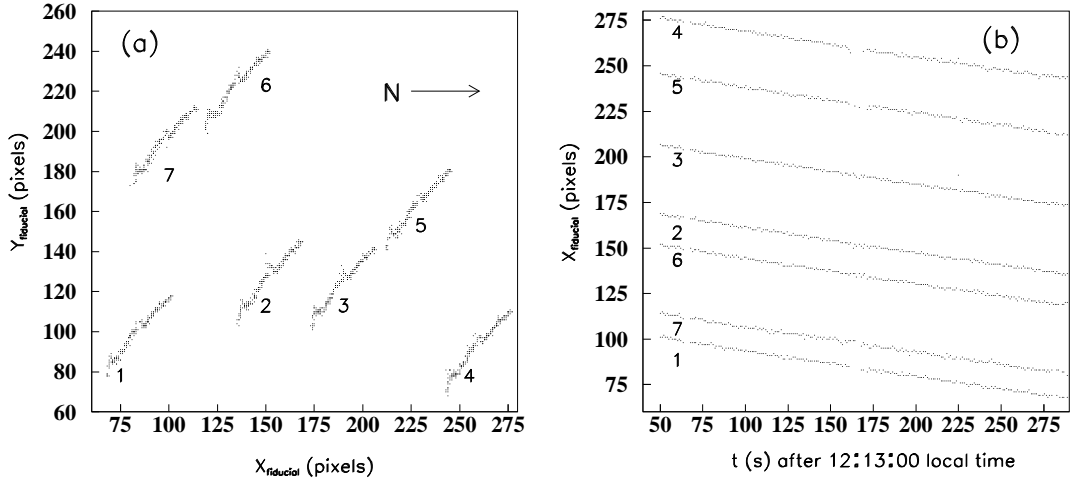}\\
  \caption{(a) Map of the successive positions in the CCD of 7 fiducial
points
on the Moon surface during the totality phase of the eclipse. The arrow
indicates the Moon North direction. (b) ``x" coordinates
of the fiducial points versus time during the eclipse.}
\label{figmap}
  \end{center}
\end{figure}

We used 7 luminous spots on the frames (small Moon
craters reflecting the ashen light) as ``fiducial" points
to reconstruct for each frame the relative position of the center of the Sun.
Although the telescope movement was set to follow the Sun, the human
activity around our apparatus caused some ground oscillations that were
 corrected using the fiducial points. For the analysis we retained
195 frames in which
all 7 points were clearly determined. Fig. \ref{figmap}a shows the
displacement of the fiducial points due to the
relative movement of the Moon with respect to the Sun. The arrow indicates
the North direction (for the Moon). In Fig. \ref{figmap}b the variation of
the ``x" coordinate with time is shown. The origin of the time axis
is at 12 hours and 13 minutes local time, about 50 seconds
before the second contact of the eclipse. Small oscillations caused by
 human activity in the vicinity of the instrument are visible; their effect
was removed in the off-line analysis.

\begin{figure}
  \begin{center}
  \includegraphics[width=9cm]{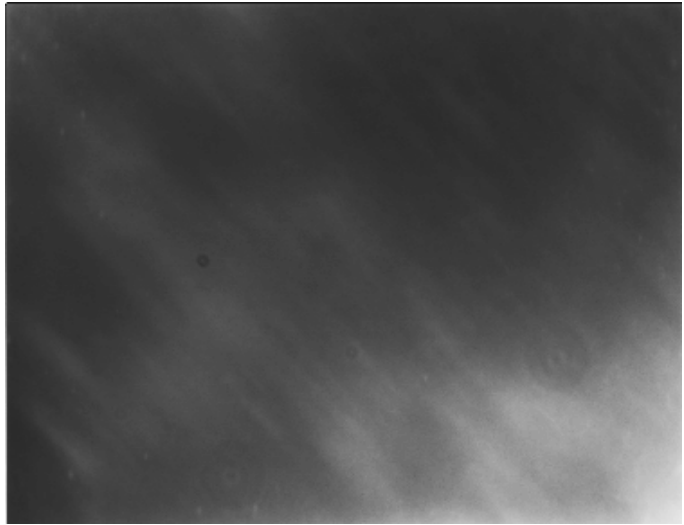}\\
  \caption{The superposition of the 195 selected frames used in the analysis.
The center of the image corresponds to the position of the center of the Sun.}
\label{sun}
  \end{center}
\end{figure}

The superposition of the selected 195 frames
is shown in Fig. \ref{sun}. All lunar landscape details are washed away; the
light gradient is due mostly to the asymmetry of the coronal light diffracted
by the Moon border and reaching the instrument. The small darker spot visible
in the picture (at the center of the image, $\sim$1.5 cm to the left) is due
to a sand grain on the
telescope window. The procedure applied to eliminate all spurious
effects on the image is discussed in Sect. \ref{ana}.

\subsection{Exposure time}
\label{timp}

The exposure time for
each frame was computed to have the maximum contrast
avoiding at the same time saturated pixels. The algorithm analyzed the
previous frame, and computed the exposure time for the next one so that the
average luminosity was in the middle of the dynamical range of the 16
bit CCD. The exposure time and other information concerning
the telescope movement were registered in the headers of each frame.

\begin{figure}
\vspace{-5cm}
  \begin{center}
  \includegraphics[width=12cm]{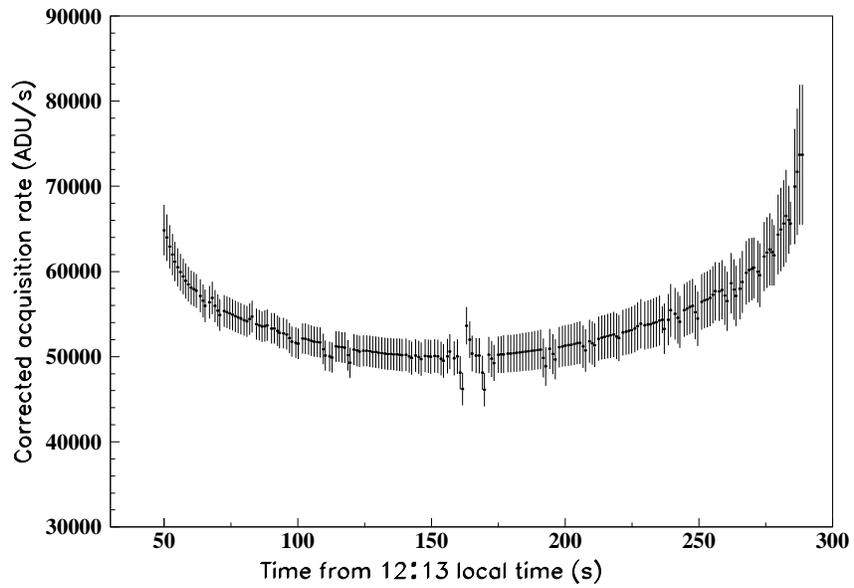}\\
 \caption{The exposure rate for each frame after the correction discussed
in the text.}
\label{corect}
 % /yp/user/popa/unix/opera_disk/bu/nottelibia/2009/corect.kumac
  \end{center}
\end{figure}

For most frames the exposure time was $0.5 \div 0.6$ seconds, as expected.
40 frames had a longer exposure, $\sim$1 s;
each of them was acquired immediately after a frame having a normal
exposure time, but with a luminosity about one half of the expected one.
This was probably caused
 by a delay in the electromechanical CCD shutter operation due to some
grains of dust that made their way into the apparatus. The exposure
times were corrected in the analysis requiring that the acquisition rate be
the average
between the acquisition rates of two adjacent frames.

Fig. \ref{corect} shows the time dependence of the frames exposure rate
in Acquisition Digital Units (ADU) per second. Larger rates at the
beginning and at the end of the totality phase are due to a larger amount
of coronal light diffracted by the Moon and reaching the telescope.
The oscillations close to the eclipse maximum correspond to
frames for which not all the 7 fiducial
points could be clearly identified (this was due to people moving
around the experiment). These frames were excluded from the final analysis.
After the time correction and the removal of ``shaken" frames, the total
exposure time during the totality phase of the eclipse is about 116 seconds.
The difference between this value and the duration of the totality is mainly
due to the time requested to read the CCD information and transfer it to the
computer.

\subsection{Calibration}
\label{cali}

As mentioned in Sect. \ref{main}, the calibration of the main system
was done the night before the eclipse using two standard luminosity stars,
SAO99215 and SAO99802. We took 20 frames, 2 s exposure each, for the
first star and 30 frames with the same exposure for the second star (which is
slightly fainter). After the removal of the dark frames the
astro-photometric measurements indicate that one ADU  corresponds to
$6.1 \pm 0.1$ photons.

\section{Monte Carlo simulations}
\label{mc}

The analysis of the data obtained by this experiment required a detailed
Monte Carlo simulation, including the neutrino production in the solar
core, its propagation, decay, and the detection by our telescope on Earth.
Such a code was developed for the analysis of the previous data collected
during the 2001 eclipse \cite{noi} and was adapted to the conditions of
the 2006
observations. The model is described in detail in ref. \cite{mc}; here
the main ideas are summarized and the parts relevant for the 2006 eclipse.

Solar neutrino production was simulated according to the ``BP2000"
Standard Solar Model (SSM) \cite{ssm} available in numerical form
in ref. \cite{ssmf}. We chose a specific reaction/decay
yielding neutrinos (both from the p-p and the CNO cycles); the
neutrino energy and the position  of its production point in the core of
the Sun were generated according to the SSM. As we are interested only in
neutrinos that could undergo radiative decays between the area of the Moon
seen by our experiment and the Earth, a decay point was generated and
the arrival direction of the decay photon was chosen. Once the geometry
of the event is known, the photon energy is computed taking into account the
Lorentz boost, and for visible photons the probability density of the angular
distribution, depending on whether the neutrino is a Dirac neutrino, left or
right-handed, or a Majorana neutrino. The number of visible photons is
about $5 \cdot 10^{-4}$ of the total spectrum; this quantity is
integrated over all directions accessed by the instrument aperture.

\begin{figure}
\vspace{-10cm}
  \begin{center}
  \includegraphics[width=14.5cm]{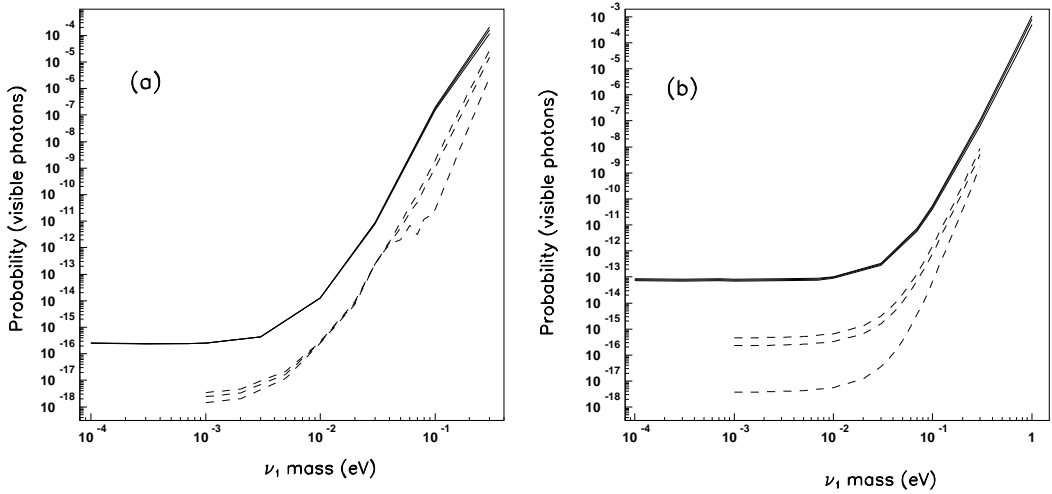}\\
  \caption{(a) The Monte Carlo probability to have a visible photon
arriving to the telescope from the $\nu_2 \rightarrow \nu_1 + \gamma$
radiative decay, versus the $\nu_1$ mass, in the conditions of the 2006
experiment (solid lines), compared to similar probabilities in the
2002 experiment (dashed lines). The dashed lines correspond from up to down
to left-handed, Majorana and  right-handed neutrinos. The differences in the
probabilities for the 2006 simulation for different $\nu_2$ polarities are
too small to be seen in the graph. (b) The same as in (a) for
the $\nu_3 \rightarrow \nu_{2,1} + \gamma$ decays.} \label{figprob}
  % /yp/user/popa/unix/opera_disk/notte2006/simnotte/probprob.kumac
  \end{center}
\end{figure}

The interpretation of solar neutrino oscillation data in the simplest two
flavour model (assuming that the electron neutrino $\nu_e$ is a superposition
of two mass eigenstates $\nu_1$ and $\nu_2$) yields a square mass difference
$\Delta m^2_{1,2} = 6 \cdot 10^{-5}$ eV$^2$ \cite{sno}. This value
 was used in simulating
$\nu_2 \rightarrow \nu_1 + \gamma$ decays. The probabilities that the
photon emitted in the space between the Moon and the
Earth is a visible photon and reaches the telescope are shown in
Fig. \ref{figprob}a, versus the $\nu_1$ mass (solid line).
For comparison, we also show the corresponding probabilities obtained in
 2002 with a different system. In
the conditions of the 2006 experiment, differences among
neutrino polarities are smaller and cannot be observed
in the scale of Fig. \ref{figprob}.

Assuming a mixing among all 3 mass eigenstates with
$\Delta m^2_{1,3} \simeq \Delta m^2_{2,3} = 2.4 \cdot 10^{-3}$ eV$^2$
(as known from  atmospheric neutrino  \cite{soudan,macro,sk} and long baseline
oscillation experiments \cite{lbl}), $\nu_3 \rightarrow \nu_2 + \gamma$
and  $\nu_3 \rightarrow \nu_1 + \gamma$ decays should behave in the
same way. The resulting probabilities
are shown in Fig. \ref{figprob}b. The notation of the lines is as
in Fig. \ref{figprob}a.

The $\nu_2 \rightarrow \nu_1 + \gamma$ decays should produce a spot of
light coincident with the center of the Sun behind the Moon disk, about 60"
large, while the signal from the $\nu_3 \rightarrow \nu_{1,2} + \gamma$
decays would consists of light rings about 20" thick, with diameters of
about 200" and 250", also centered in the Sun.

\section{Data analysis}
\label{ana}

The search for $\nu_2 \rightarrow \nu_1 + \gamma$ and
$\nu_3 \rightarrow \nu_{1,2} + \gamma$ signatures in the frames recorded
during the 2006 total solar eclipse is based on the wavelet decomposition
of the compound image, shown in Fig. \ref{sun}. We recall that this image is
a superposition of 195 selected 16 bit frames recorded during the TSE
and aligned with respect to the center of the Sun position behind the Moon.
As in the case of the 2002 eclipse \cite{noi} we chose the Haar
wavelet basis \cite{har}. The $n$-order term of the decomposition is obtained
by dividing the $N \times N$ pixels$^2$ image in squares of
$N/2^n \times N/2^n$ pixels$^2$ and averaging the luminosity in each
square; the averages are then subtracted and the resulting image,
the $n$-order
residual, can be used to obtain the $(n+1)$-order term. Each
decomposition term is an image that enhances the features
 on the corresponding scale, while the residuals yield information
for smaller dimension scales.

The decay signal is searched for by averaging the luminosity of the images
over concentric ``rings" centered in the Sun. As the
wavelet analysis requires a dyadic dimension (the number of
pixels has to be an integer power of 2), we considered four pixels adjacent
to the image center
as ``central" and then averaged the obtained luminosity profiles.

\begin{figure}
\vspace{-6cm}
  \begin{center}
  \includegraphics[width=12cm]{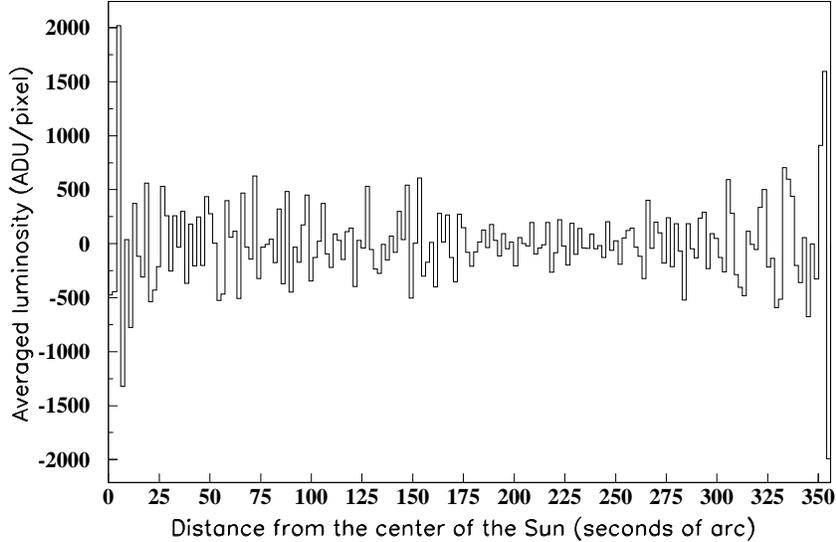}\\
 \caption{Luminosity profile of the $7^{th}$ residual from the
wavelet decomposition of the central part of the Moon image in
Fig. \ref{sun}.}
\label{wave}
 % /yp/user/popa/unix/opera_disk/bu/nottelibia/2009/profil1.kumac
  \end{center}
\end{figure}

The image in Fig. \ref{sun} is $256 \times 256$ pixels$^2$; we used up to
 the $7^{th}$ order in the wavelet decomposition. Fig. \ref{wave} shows
the luminosity distribution for the $7^{th}$
order residual, that may yield information on the smaller
scale effects in the original image. The profile is consistent with
statistical noise; the larger amplitudes near the center of the Sun
and near the edges of the image are due to a smaller number of
pixels available for the averaging.

None of the image decomposition terms or residuals shows structures as
would be expected from solar neutrino radiative decays; thus
 a possible signal cannot be larger than the statistical fluctuations.

\section{Results and discussions}
\label{res}

The expected scale of the decay signal (few tens of arcseconds) suggests
that the wavelet term most sensitive to
$\nu_2 \rightarrow \nu_1 + \gamma$ and $\nu_3 \rightarrow \nu_{1,2} + \gamma$
decays should be the $5^{th}$ order, with a typical scale of about 16".
The number of  visible photons originating from solar neutrino radiative
decays that could be recorded by the telescope CCD may be computed as
\begin{equation}
N_\gamma = P \Phi_{(2,3)} S_M t_{obs} \left(1 - e^{\frac{\langle t_{ME} \rangle}{\tau}} \right) e^{- \frac{t_{SM}}{\tau_{(2,3)}}},
\label{final}
\end{equation}
where $P$ is the mass-dependent probability shown in Figs.
\ref{figprob}a,b and $\Phi_{(2,3)}$ represents the flux of $\nu_2$
or $\nu_3$ solar neutrinos at the Earth.
\begin{equation}
 \begin{array}{ll}
 \Phi_2 = \Phi_\nu \sin^2 \theta_{12} \\
 \Phi_3 = \Phi_\nu \sin^2 \theta_{13},
 \end{array}
 \label{flux}
 \end{equation}
 where $\Phi_\nu  \simeq 7 \cdot 10^{10}$ cm$^{-2}$s$^{-1}$ is the expected
solar neutrino flux at the Earth (neglecting oscillations) and $\theta_{12}$
is the mixing angle in the two flavour approximation. For 3 neutrino flavours
the mixing angle $\theta_{13}$ in uncertain; in our calculations
we used $ \sin^2 \theta_{13}= 0.1$.
If one used 0.06 as the 95\% C.L. quoted by SNO \cite{sno}, there would be
only a slight reduction of sensitivity, while the reduction would be
considerable in the case of $ \sin^2 \theta_{13}= 0.02$ \cite{sno,pdg}.
In Eq. \ref{final} $S_M$ is
the area on the Moon surface covered by our observations, $t_{obs}$ is the
total acquisition time, $\langle t_{ME} \rangle$ is the average travel time
of the neutrinos in the observational cone from the Moon to the Earth
(assuming that the decay point is uniformly distributed along that
distance $\langle t_{ME} \rangle$ is about one third of the flying
time \cite{noi}), $t_{SM}$ is the flight time from the Sun to the
Moon and $\tau_{(2,3)}$ the lifetime of $\nu_2$ and  $\nu_3$ neutrino
mass higher states. All time variables in Eq. \ref{final} are defined in the
laboratory frame of reference.

\begin{figure}
\vspace{-9cm}
  \begin{center}
  \includegraphics[width=15cm]{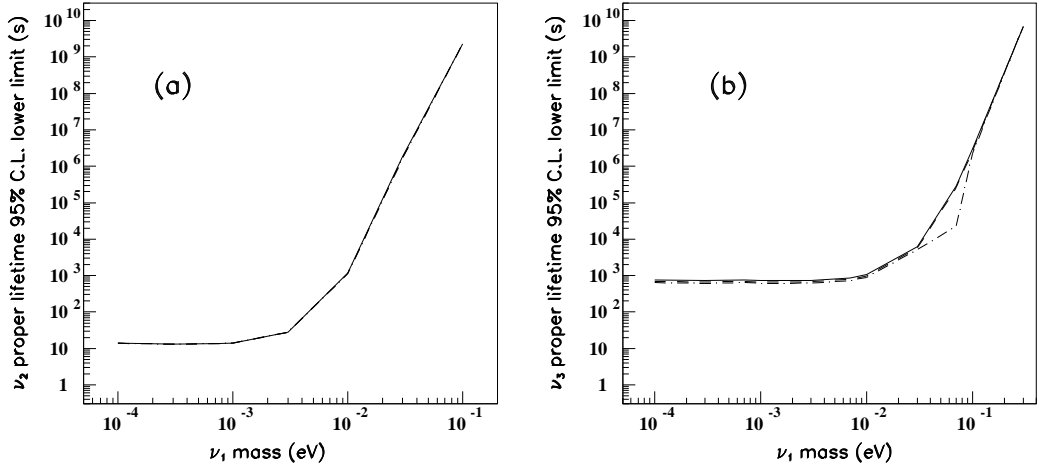}\\
  \caption{(a) 95\% C.L. lower limits for the $\nu_2$ proper lifetime.
 The differences between different polarization states cannot be seen at
this scale. (b) 95\% C.L. lower limits for the $\nu_3$  proper lifetime,
assuming $\sin^2 \theta_{13} = 0.1$. The lines correspond from up to
down to left-handed, Majorana and right-handed neutrinos.}
\label{lim}
  % /yp/user/popa/unix/opera_disk/noiembrie2009/2009/finalfinal.kumac
  \end{center}
\end{figure}

No structure compatible with $\nu_2$ or $\nu_3$ radiative decays
was found in our analysis; lower limits for the lifetimes of the heavy
neutrino components were obtained. The number of photons
produced through radiative decays between the Moon and the Earth reaching our
detector is $N_\gamma \leq 3\sigma_{T5}$ (95\% C.L.), where $\sigma_{T5}$ is
the standard deviation of the luminosity of the $5^{th}$ term in the wavelet
decomposition of the data. The 95\% C.L. lower limits for the $\nu_2$ proper
lifetimes are shown in Fig. \ref{lim}a. Although they were computed assuming
three possible $\nu_2$ polarizations, the results are so close that cannot
be separated on the graph. For the $\nu_3$ proper lifetimes, the
95\% C.L. upper limits, computed assuming $\sin^2 \theta_{13} = 0.1$, are
shown in Fig. \ref{lim}b. The solid line corresponds to left-handed Dirac
neutrinos, the dash-dotted line to Majorana neutrinos and the dashed line
to right-handed Dirac neutrinos.

\section{Conclusions}
\label{conc}

The analysis of 195 frames recorded in occasion of the $26^{th}$ of March
2006 total solar eclipse in the Libyan Sahara desert did not evidence any
signal compatible with the Monte Carlo predictions for the radiative decays
of the heavier components of solar neutrinos \cite{mc}.

For $\nu_2 \rightarrow \nu_1 + \gamma$ radiative decay the 95\% C.L.
lower lifetime limits are in the range $10 \div 10^9$ s, for neutrino masses
$10^{-4} < m_{\nu1} < 0.1$ eV. These limits represent an improvement of
about $2 \div 3$
orders of magnitude in the lower neutrino mass region  with respect to our
previous results \cite{noi}. Similar improvements were obtained for the
 $\nu_3 \rightarrow \nu_{2,1} + \gamma$, but the limits are tentative since
the mixing angle $\theta_{13}$ is still uncertain.

 The limits mentioned above are obtained from the fifth order wavelet
term of the summed image of Fig. \ref{sun}. The detection of the
ashen light combined with the negative result prove that the searched
signal should be fainter than the ashen light itself; we thus can state that
the limits presented in this paper are the best obtainable using this
technique, since there is no way to avoid the ashen light background.

\section*{Acknowledgments}

{\small We would like to acknowledge many colleagues for useful comments and
discussions. The experiment was funded by the University and INFN Section
of Bologna. We acknowledge also the support from the Italian Institute of
Culture of Tripoli. The analysis was partially funded under CNCSIS Contract
539/2009.
We are grateful to the organizers of the SPSE 2006 event for their efforts
that allowed to perform our experiment. Special thanks are due to
the Winzrik Group and to the Libyan Air Force for their assistance.}

\end{document}